# Investigating the Origins of Spiral Structure in Disk Galaxies through a Multiwavelength Study

Ryan Miller[1,2], Daniel Kennefick[1,2], Julia Kennefick[1,2], Mohamed Shameer Abdeen[1,2], Erik Monson[1,2],

Rafael T Eufrasio[1], Douglas W Shields[1,2], and Benjamin L Davis[3]

[1] Department of Physics, University of Arkansas, 226 Physics Building, 825 West Dickson Street, Fayetteville, AR 72701, USA; ryan.mlr83@gmail.com

[2] Arkansas Center for Space and Planetary Sciences, University of Arkansas, 346 1/2 North Arkansas Avenue, Fayetteville, AR 72701, USA

[3] Centre for Astrophysics and Supercomputing, Swinburne University of Technology, Hawthorn, Victoria 3122, Australia

## Abstract

The density-wave theory of spiral structure proposes that star formation occurs in or near a spiral-shaped region of higher density that rotates rigidly within the galactic disk at a fixed pattern speed. In most interpretations of this theory, newborn stars move downstream of this position as they come into view, forming a downstream spiral which is tighter, with a smaller pitch angle than that of the density wave itself. Rival theories, including theories which see spiral arms as essentially transient structures, may demand that pitch angle should not depend on wavelength. We measure the pitch angle of a large sample of galaxies at several wavelengths associated with star formation or very young stars (8.0 $\mu$m, H-$\alpha$ line and 151 nm in the far-UV) and show that they all have the same pitch angle, which is larger than the pitch angle measured for the same galaxies at optical and near-infrared wavelengths. Our measurements in the *B* band and at 3.6 $\mu$m have unambiguously tighter spirals than the star-forming wavelengths. In addition we have measured in the *u* band, which seems to fall midway between these two extremes. Thus, our results are consistent with a region of enhanced stellar light situated downstream of a star-forming region.

## 1. Introduction

The density-wave theory has dominated the interpretation of spiral arms in disk galaxies since the mid-sixties (Lin & Shu 1964; Bertin & Lin 1995; Shu 2016). The original theory, with its disk-spanning standing-wave pattern created by resonant modes, has been challenged by numerical simulations which suggest that the waves should be subject to damping which probably prevents long-lasting modes from generating semi-permanent spiral arms.

An alternative mechanism for the production of spiral density waves, known as swing amplification (introduced by Goldreich & Lynden-Bell 1965 and Julian & Toomre 1966), proposes that small local disturbances can be amplified to create transient patterns. Eventually detailed simulations provided strong evidence that density waves are probably incapable of producing the kind of semi-permanent patterns called for by the original theory and that some form of amplification along the lines of swing amplification must play an important role in periodically causing transient spiral patterns to recur (Sellwood & Carlberg 1984). Currently, the theoretical situation is such that quite diverse views co-exist and there is no consensus that a single mechanism is responsible for all of the observed spiral patterns in galaxies. While some experts insist that spiral patterns typically last only a galactic rotation or two (Sellwood & Binney 2002; Grand et al. 2012), other theorists argue that swing amplification can give rise to superposed modes of the system which can last for up to ten rotation periods (Sellwood & Carlberg 2011). This issue is relevant to the current paper because it has been argued that failure to find consistent downstream displacements of observational tracers of star formation and stars of different ages is an argument against the reality of long-lasting spiral arms (Foyle et al. 2011). By contrast, in previous work (Pour-Imani et al. 2016) we confirmed a key prediction of the density-wave theory, that spiral-arm pitch angle varies with observation wavelength. This is in contrast to the predictions of rival theories, such as the Manifold theory (for a discussion of this theory test see Athanassoula et al. 2010).

The density-wave theory predicts that the density-wave gives rise (through compression of clouds approaching and passing through it) to a star-forming region and that newly born stars will move downstream of this star-forming region before they are observed. How this affects pitch angle depends crucially on the existence of a corotation radius, a point on the disk at which the rotational speed of stars equals the rotational speed of the fixed spiral pattern itself (the density wave)(Peterken et al. 2019). Inside the corotation radius, stars move faster than the pattern speed and downstream means "in advance of the pattern." Outside the corotation radius stars move slower than the pattern speed and downstream means "falling behind the pattern." Thus, newly born stars should form a spiral arm that is tighter (with a smaller pitch angle) than the spiral density wave itself, as shown in Figure 1. In an earlier paper (Pour-Imani et al. 2016) we showed that for a sample of 41 galaxies there is a clear difference between the pitch angle of two wavelengths associated with stellar light, the *B* band and 3.6 $\mu$m and two wavelengths associated with star formation, 8.0 $\mu$m and 151 nm in the far ultraviolet (FUV). The stellar pitch angles are uniformly smaller (tighter spiral arms) than the star formation pitch angles. In this paper, we look at another wavelength associated with star formation, the H-$\alpha$ line. In confirmation of the earlier result, the pitch angles for this wavelength agree well with those previously measured for 8.0 $\mu$m and the FUV. In addition, we have added pitch angles measured in the *u* band. This band lies midway between the FUV and the *B* band, which, though close in wavelength, disagree in pitch angle. Not surprisingly, we find evidence that the pitch angle associated with the $u$-band appears to lie between these other two, suggesting that there are some stars that do not live long enough to move from the star-forming region (seen in the FUV), stars which move a short distance away ($u$ band), and stars that live long enough to move clearly away from the star-forming region ($B$ band).

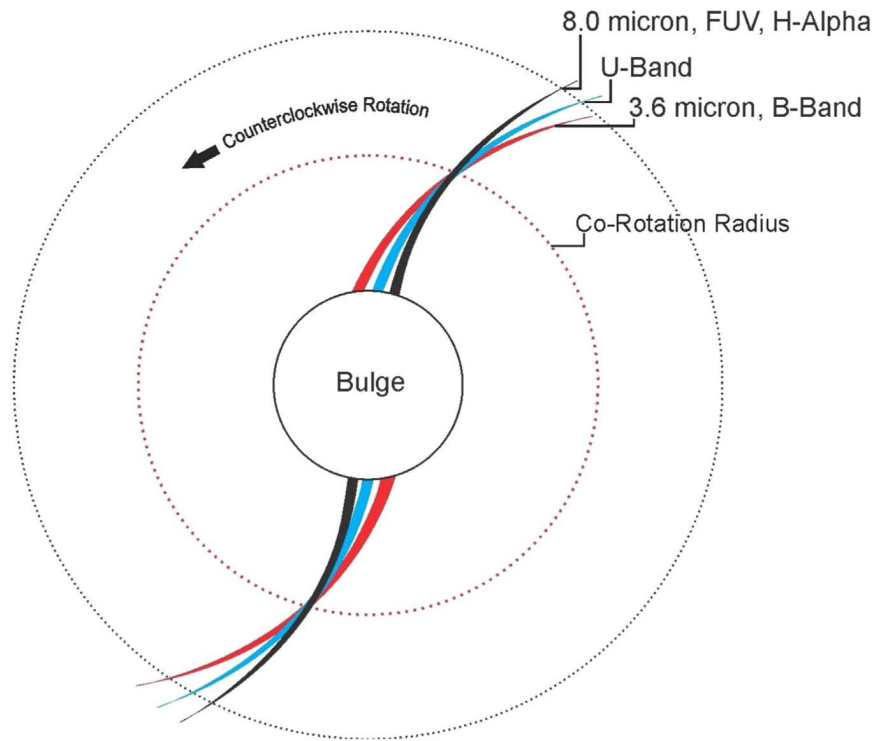

**Figure 1.** Illustration of spiral-arm structure based upon pitch angle measurements at multiple wavelengths in this paper. We observe tighter arms (lower pitch angle) for 3.6 $\mu$m and the $B$ band, looser arms (higher pitch angle) for the $u$ band and even looser arms for 8.0 $\mu$m, FUV, and H-$\alpha$ images. Thus, the star-forming arm is upstream from the blue arm, which is in turn upstream from the red arm. This conforms with the density-wave theory's prediction of variation in pitch angle with the image wavelength of light (Pour-Imani et al. 2016).

The existence of a color gradient from blue to red downstream from the star-forming region is not, however, the only prediction of the modal density wave theory. Because the process of gravitational collapse itself takes time it is likely that the star-forming region is itself downstream from the actual density wave. Thus, upstream from the star-forming region there may be an opposite color gradient, which goes from red, old disk stars compressed by the density wave, to blue young stars found in the star-forming region. As we have seen, upstream spirals have looser pitch angles than downstream spirals, therefore the density-wave spiral itself has a looser pitch angle than the star formation spiral and therefore the redder arm is looser than the blue arm, which is the opposite for the gradient on the other side of the star-forming region described above. One has to be careful here though. Although clearly one would expect a looser red spiral upstream from a tighter blue-spiral, there might not be a continuous gradient from blue to red, since the red spiral created by the density wave compressing old disk stars close together affects all old disk stars equally, regardless of their particular age or color. Furthermore, when considering either of these two color gradients, one must remember that extinction may complicate matters by obscuring some wavelengths more than others.

In considering these two possible color gradients predicted by density wave theory, a blue to red gradient downstream from the star-forming region and a red-to-blue gradient upsteam from it, our results come down in favor of the former. An important point is that the $B$-band and 3.6 $\mu$m pitch angles agree reasonably well with each other. If anything we see the 3.6 $\mu$m pitch angle as being even a little tighter than the $B$-band one. In short, there may be a steady gradient from blue to red of tightening pitch angles, moving through FUV, the $u$ band, and the $B$ band to 3.6 $\mu$m. However, some other groups report a different result, in line with the expectation that in the near-infrared (NIR) one views the old red disk stars compressed together by the density wave itself (Grosbol & Patsis 1998; Martínez-García 2012; Martínez-García et al. 2014). This red-to-blue gradient is just as much a prediction of the density wave theory as the blue-to-red gradient reported by us, but there is obviously a question as to which one is really observable! This is a difficult question to answer, because the differences between $B$-band pitch angles and 3.6 $\mu$m pitch angles are typically small. We shall discuss this issue in more detail below, but the overall picture right now is that the result reported by some groups (Seigar et al. 2006; Davis et al. 2012), that the difference between pitch angles measured in these two bands is very small at best, may be the most that can be said right now. It has been proposed to us that an interpretation is possible that unites both of these scenarios. If it should happen that the star-forming region is upsteam from the density wave, then it could follow that the density-wave compression red spiral would exist downstream and roughly overlap with the blue-to-red spiral reported by us. Presumably, this could happen if the approach to the region of maxiumum density-wave compression was enough to kick start collapse in gas clouds before they actually crossed the maximum of the spiral potential. Although such a possibility cannot be ruled out, it does not seem to be reflected in most simulations of galactic density waves. Nevertheless, we illustrate the scenario in Figure 2.

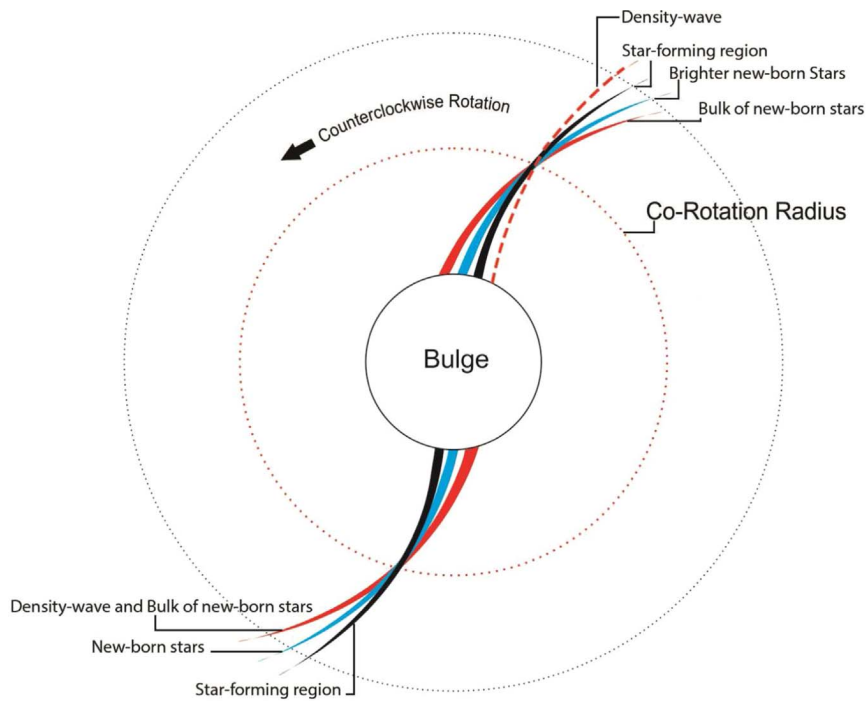

**Figure 2.** Two different scenarios are illustrated here to explain the results shown in Figure 1, in which wavelengths associated with star formation are seen upstream from bright young stars, which in turn are found in the vicinity of redder starlight. In the first scenario, illustrated in the upper arm, the density wave kickstarts star formation and both of these regions are upstream of the newborn stars, which are spread out downstream. In this scenario the 8 $\mu$m, FUV, and H-$\alpha$ images are tracers for the star-forming arm, and B and NIR wavebands show the location of newly born stars downstream from that. In the second scenario, illustrated in the lower arm, star formation is initiated as gas clouds approach the location of the spiral density wave. Bright newborn stars are born and first seen between the position of the star-forming region and the density wave. In this scenario, newly born stars seen downstream from the star-forming region will be found close to the position of the density wave. The density wave itself is visible because it compresses the older redder background disk stars, and makes them more visible. In this scenario the two different gradients discussed in the text might coincidentally be superimposed upon each other, so that the red spiral arm created by density-wave compression is in more or less the same location as the spiral arm created by newly born stars. However, it must be noted that there is no great theoretical warrant for this scenario.

## 2. Data

Our sample of 29 galaxies is drawn from the *Spitzer* Infrared Nearby Galaxies Survey, which consists of imaging from the Infrared Array Camera (Fazio et al. 2004). The sample is drawn from the one found in Pour-Imani et al. (2016) by selecting those objects for which images in the *u* band and the H-$\alpha$ line are available. Thus the sample in this paper selects those galaxies with imaging at 3.6 $\mu$m that had available optical imaging in the *B* band (445 nm) and ultraviolet imaging in the *u* band (355 nm) as found in the NASA/IPAC Extragalactic Database (NED; see Table 1 for *B* band and *u* band image sources). Twenty-eight of these galaxies also have available ultraviolet imaging from archived *Galaxy Evolution Explorer* (*GALEX*) data in the FUV 1516 Å and 14 galaxies with narrowband H-$\alpha$ imaging as indicated in Table 1. Our data cover a representative range of galaxy environment (i.e., isolated galaxies, interacting galaxies, group members, and cluster members). The hierarchy reported by NED shows nine galaxies are isolated, 15 galaxies are group galaxies, and five galaxies are in pairs.

## 3. Results

Spiral-arm pitch angles of our sample galaxies were measured in six wavebands (depending on availability) using the 2DFFT code, whose use is described in detail in Davis et al. (2012; for the source code see Davis et al. 2016). As its name implies it is a two-dimensional fast Fourier tranform code that breaks down galaxy images into a superposition of spirals of different pitch angles and numbers of arms, identifying the strongest mode to measure the pitch angle. The 2DFFT is taken over an annulus of the galaxy defined by an inner and outer radius. The user chooses the outer radius to coincide with the edge of the galaxy's visible disk. The code then runs for every possible inner radius from the center of the galaxy (which is user defined) and plots the measured pitch angle for each of these defined annuli as a function of inner radius. The user then selects a "stable region" of inner radius over which (for a given number of arms, usually the strongest Fourier mode) the pitch angle varies as little as possible. The measured pitch angle is the average over this range of inner radii, with the variance forming the basis of the estimated error (weighted to penalize galaxies with short stable regions). Measurements were also made using a completely independent code, called Spirality, described in Shields et al. (2015a). Spirality uses an orthogonal spiral coordinate system to find the pitch angle of the coordinate arm that has the maximum brightness along it when superimposed over the galaxy. Only the 2DFFT pitch angles are reported in this paper, the Spirality values being measured merely as a check on our results. As an example, in one case a

Table 1
Sample

| Galaxy Name (1) | Type (2) | P (3.6 μm) (3) | P (B band) (4) | P (u band) (5) | P (8.0 μm) (6) | P (H-α) (7) | P (FUV) (8) | Inclination (9) | Image Source (10) |
|---|---|---|---|---|---|---|---|---|---|
| NGC 0628 | Sac | 9.58 ± 0.60 | 9.20 ± 0.83 | 19.76 ± 2.08 | 20.60 ± 2.28 | ... | 21.43 ± 1.42 | 32.3 | IRAC, NOT, *GALEX* |
| NGC 0925 | SABd | 4.45 ± 0.65 | 7.51 ± 3.81 | 52.90 ± 8.10 | 20.10 ± 4.69 | 25.16 ± 3.40 | 29.68 ± 3.75 | 53.1 | IRAC, PAL, KPNO2, *GALEX* |
| NGC 1097 | SBb | 6.84 ± 0.21 | 7.54 ± 3.49 | 11.62 ± 2.70 | 9.50 ± 1.28 | 12.10 ± 3.10 | 16.25 ± 2.3 | 46.2 | IRAC, LCO, duPont, 2MASS, *GALEX* |
| NGC 1566 | SABbc | 15.29 ± 2.37 | 31.20 ± 4.80 | 35.50 ± 1.30 | 44.13 ± 11.94 | ... | 45.80 ± 2.97 | 34.3 | IRAC, KPNO, duPont, *GALEX* |
| NGC 2403 | SABc | 12.50 ± 1.62 | 19.35 ± 1.57 | 20.33 ± 2.10 | 28.52 ± 6.73 | 28.07 ± 2.80 | 23.54 ± 0.78 | 49.8 | IRAC, LCO, 2MASS, Palomar, *GALEX* |
| NGC 2841 | SAb | 16.13 ± 1.63 | 18.77 ± 1.66 | 18.68 ± 3.40 | 22.25 ± 2.42 | ... | 23.26 ± 2.31 | 62.5 | IRAC, LOWE, Palomar, *GALEX* |
| NGC 2976 | Sac | 4.14 ± 0.34 | 5.13 ± 0.43 | 9.80 ± 1.30 | 8.36 ± 0.40 | ... | 10.68 ± 1 | 54.8 | IRAC, KPNO, SDSS, *GALEX* |
| NGC 3031 | SAab | 15.63 ± 6.99 | 16.19 ± 1.23 | 19.70 ± 1.50 | 20.54 ± 2.21 | ... | 20.14 ± 1.90 | 55.3 | IRAC, JKY, *GALEX* |
| NGC 3184 | SABcd | 11.92 ± 1.77 | 18.30 ± 3.45 | 11.27 ± 4.50 | 23.40 ± 3.27 | 23.45 ± 1.90 | 26.75 ± 0.55 | 20.3 | IRAC, KPNO, NOT, 2MASS, *GALEX* |
| NGC 3190 | SAap | 16.39 ± 2.15 | 17.67 ± 2.34 | 26.75 ± 7.02 | 18.35 ± 4.43 | ... | ... | 43.1 | IRAC, CTIO, SWIFT |
| NGC 3198 | SBc | 15.97 ± 1.38 | 18.95 ± 2.69 | 20.46 ± 4.10 | 20.59 ± 5.95 | ... | 23.98 ± 1.84 | 71.5 | IRAC, CTIO, SDSS, *GALEX* |
| NGC 3351 | SBb | 4.60 ± 1.92 | 16.41 ± 1.93 | 17.45 ± 1.60 | 22.21 ± 6.96 | 20.89 ± 5.10 | 27.17 ± 2.11 | 41.6 | IRAC, CTIO, SDSS, KPNO2, *GALEX* |
| NGC 3521 | SABbc | 16.74 ± 1.32 | 19.28 ± 1.92 | 21.8 ± 1.90 | 21.48 ± 2.19 | ... | 24.81 ± 2.04 | 66.1 | IRAC, CTIO, SDSS, *GALEX* |
| NGC 3621 | SAd | 17.22 ± 3.37 | 18.43 ± 3.12 | 21.30 ± 2.98 | 20.81 ± 2.72 | ... | 20.34 ± 1.98 | 49.3 | IRAC, ESO, *GALEX* |
| NGC 3627 | SABb | 11.71 ± 0.78 | 16.97 ± 1.54 | 53.10 ± 4.20 | 18.59 ± 2.85 | ... | 40.29 ± 1.60 | 61.8 | IRAC, KPNO, SDSS, *GALEX* |
| NGC 3938 | SAc | 11.46 ± 2.32 | 12.22 ± 1.94 | 23.50 ± 2.59 | 19.34 ± 3.80 | 23.39 ± 3.30 | 21.45 ± 1.87 | 22.7 | IRAC, LOWE, 2MASS, *GALEX* |
| NGC 4254 | SAc | 28.40 ± 4.04 | 30.01 ± 4.36 | 39.32 ± 6.92 | 32.8 ± 1.45 | 33.40 ± 4.50 | 38.66 ± 3.91 | 38.3 | IRAC, INT, 2MASS, *GALEX* |
| NGC 4321 | SABbc | 18.60 ± 1.69 | 15.06 ± 1.20 | 23.17 ± 3.10 | 24.46 ± 3.76 | 21.18 ± 2.50 | 28.49 ± 1.26 | 33.2 | IRAC, KPNO2, 2MASS, *GALEX* |
| NGC 4450 | SAab | 12.59 ± 2.63 | 16.62 ± 1.45 | 21 ± 2.00 | 21.2 ± 3.87 | ... | 22.99 ± 5.43 | 49.1 | IRAC, LOWE, *GALEX* |
| NGC 4536 | SABbc | 17.3 ± 1.75 | 33.74 ± 4.8 | 52.30 ± 1.90 | 52.22 ± 2.42 | 52.37 ± 4.80 | 55.59 ± 2.75 | 59.1 | IRAC, KP, SDSS, KPNO2, *GALEX* |
| NGC 4569 | SABab | 8.64 ± 0.52 | 19.05 ± 2.42 | 22.21 ± 3.6 | 38.55 ± 6.44 | ... | 42.11 ± 5.80 | 63.2 | IRAC, PAL, SDSS, *GALEX* |
| NGC 4579 | SABb | 11.44 ± 0.57 | 13.80 ± 3.25 | 14.30 ± 1.60 | 30.73 ± 4.73 | 32.74 ± 4.10 | 33.98 ± 3.69 | 43.4 | IRAC, KPNO, SDSS, KPNO2, SWIFT |
| NGC 4725 | SABab | 3.04 ± 1.83 | 7.40 ± 0.35 | 9.49 ± 2.50 | 10.80 ± 1.04 | 15.59 ± 4.90 | 13.6 ± 1.92 | 48.2 | IRAC, KPNO4, SDSS, KPNO2, *GALEX* |
| NGC 4736 | SAab | 8.19 ± 2.94 | 8.41 ± 1.34 | 16.31 ± 1.29 | 14.09 ± 5.11 | ... | 14.98 ± 2.31 | 33.1 | IRAC, PAL, *GALEX* |
| NGC 5055 | SAbc | 16.35 ± 1.78 | 19.31 ± 1.63 | 19.10 ± 4.80 | 20.63 ± 2.11 | 21.22 ± 2.21 | 20.29 ± 5.87 | 55.9 | IRAC, PAL, 2MASS, *GALEX* |
| NGC 5474 | SAcd | 12.11 ± 1.15 | 13.84 ± 6.22 | 18.62 ± 6.57 | 19.12 ± 3.22 | ... | 19.91 ± 2.65 | 46.2 | IRAC, JKY4034, *GALEX* |
| NGC 5713 | SABb | 12.20 ± 0.32 | 18.76 ± 3.10 | 20.81 ± 3.80 | 34.79 ± 5.01 | 30.37 ± 4.10 | 27.40 ± 1.20 | 29.7 | IRAC, CTIO, 2MASS, *GALEX* |
| NGC 7331 | Sab | 17.13 ± 2.63 | 20.10 ± 1.85 | 19 ± 1.70 | 21.65 ± 2.15 | ... | 22.54 ± 2.41 | 67.6 | IRAC, KPNO, *GALEX* |
| NGC 7793 | SAd | 10.98 ± 1.60 | 12.16 ± 2.10 | 14.47 ± 3.30 | 16.34 ± 5.47 | 17.80 ± 3.90 | 16.89 ± 1.87 | 53.9 | IRAC, ESO, CTIO, 2MASS, *GALEX* |

**Note.** Columns: (1) galaxy name; (2) Hubble morphological type; (3) pitch angle in degrees for 3.6 μm; (4) pitch angle in degrees for *B*-band 445 nm; (5) pitch angle in degrees for *u*-band 355 nm; (6) pitch angle in degrees for 8.0 μm; (7) pitch angle in degrees for H-α; (8) pitch angle in degrees for FUV 151 nm; (9) inclination angle in degrees for 3.6 μm; (10) telescope/literature source of imaging.

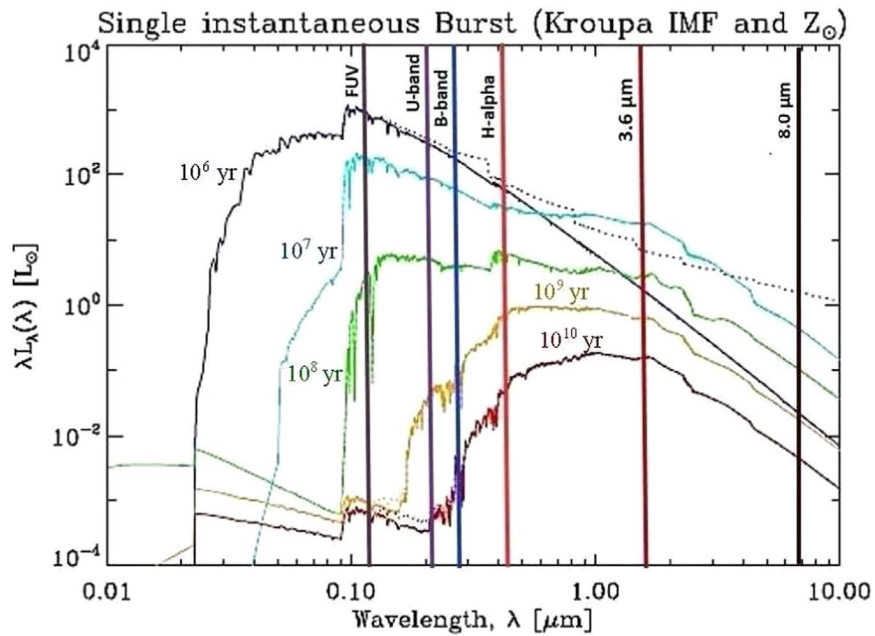

**Figure 3.** Spectral evolution for a single burst of star formation, for a Kroupa IMF and solar metallicity, $Z_\odot = 0.02$. The total initial stellar mass is normalized to $1\,M_\odot$. The solid lines show photospheric emission from the stellar population. Spectra are color-coded, with respective ages shown in the upper left corner. Models with nebular continuum emission are shown with the same color-coding. Only the youngest age shows significant difference due to its large ionizing photon rate. The figure is from Eufrasio (2015).

disagreement between Spirality and 2DFFT was discovered to be because of a poor choice of stable region, which left 2DFFT measuring only a small proportion of the galaxy disk. The measurement with 2DFFT was then redone so that it measured the same part of the disk that Spirality was sensitive to, and this resulted in the two codes reaching better agreement. In all other cases they agreed without the need to modify 2DFFT's result.

Both algorithms presume the existence of logarithmic spirals that span the disk. Therefore, it is probably true that our methods work best with grand-design spirals, although it is not at all necessary that the spirals be two-armed. Since both methods work better with a long spiral arm to measure, bars are not helpful and it is worth noting that three of the galaxies in our sample have measured pitch angles that disagree noticeably between certain wavebands and are suspiciously far from the line of equality in some panels of Figure 4. Each of these galaxies is barred and it is possible that our measurements are suspect in these cases. They have been left in so as not to cherry-pick the sample, but it is worth noting that some galaxies, specifically non-barred grand-design spirals, are probably more amenable to our methods than others. Finally, it is worth commenting on disk inclination. An important step in the process of both algorithms is "de-projecting" the image, by elongating along one axis to make it as close to circular as possible. Thus, we estimate the galaxy's inclination using IRAF's ellipse function. Our values are given in the table for each galaxy. Our results are in reasonably good agreement with those given in the S4G survey (Salo et al. 2015).

The $B$-band images are strongly sensitive to newly born stars that have emerged from their stellar nurseries. For NIR images at 3.6 $\mu$m it is expected that older stars would contribute much of the light, but we must also keep in mind that stellar evolution models suggest that newborn stars also produce a lot of red light. By contrast, the ultraviolet images taken by *GALEX* at 1516 Å are sensitive to the brightest O-type stars with the shortest lives, visible while still in or close to the star-forming region (see Figure 3). The $u$-band (355 nm) images are also expected to trace newborn stars. We might expect images from this waveband to lie between the 1516 Å and the $B$-band measurements. In other words, the $u$ band may trace stars that are short-lived, but still have time to move some distance from where they are born.

Our results show that the pitch angles of 28 galaxies in the FUV(1516 Å) are bigger than the $u$-band (355 nm) pitch angles (corresponding to looser arms) in most of the cases we examine. However, as can be seen from the adjoining histogram (Figures 4 and 5), the typical difference in pitch angle is less than the average measurement error. Therefore, we cannot clearly distinguish between these two wavelengths. Similarly, as seen in the second panel of Figure 4, the pitch angles of 29 galaxies (the entire sample) in the $u$ band (355 nm) are larger than those measured from the $B$-band (445 nm) image. Again, the difference is typically less than the measurement error, which is unsurprising since they are so close, we cannot clearly distinguish between these two wavelengths. But note that we can clearly distinguish between the FUV and $B$ band (third panel of Figure 4, reprinted from Pour-Imani et al. 2016), as argued in our previous work (Pour-Imani et al. 2016).

It seems clear that if FUV is decisively different from the $B$ band, but the $u$ band is not clearly separated from either, than logically the $u$ band tends to lie between these two (see Figure 6). This makes sense if we imagine that what we are seeing is progressively longer-lived stars moving downstream from where they were formed. Finally, the last panel shows that the 3.6 $\mu$m and $u$-band images clearly disagree in pitch angle. The 3.6 $\mu$m images have consistently tighter pitch angles, with the histogram showing that they typically differ by 4° or more. Again, this seems to add weight to our argument that we are seeing a gradual decrease in pitch angle from blue to red, beginning with the FUV, then with the $u$ band, then the $B$ band and finally the NIR. But note that the bulk of the change takes place within the *uv*. Within the entirety of the optical and NIR, there is no change in pitch angle greater (on average) than our typical measurement error. Figure 6 shows the evolution of pitch angle from FUV (151 nm) to 3.6 μm for our samples.

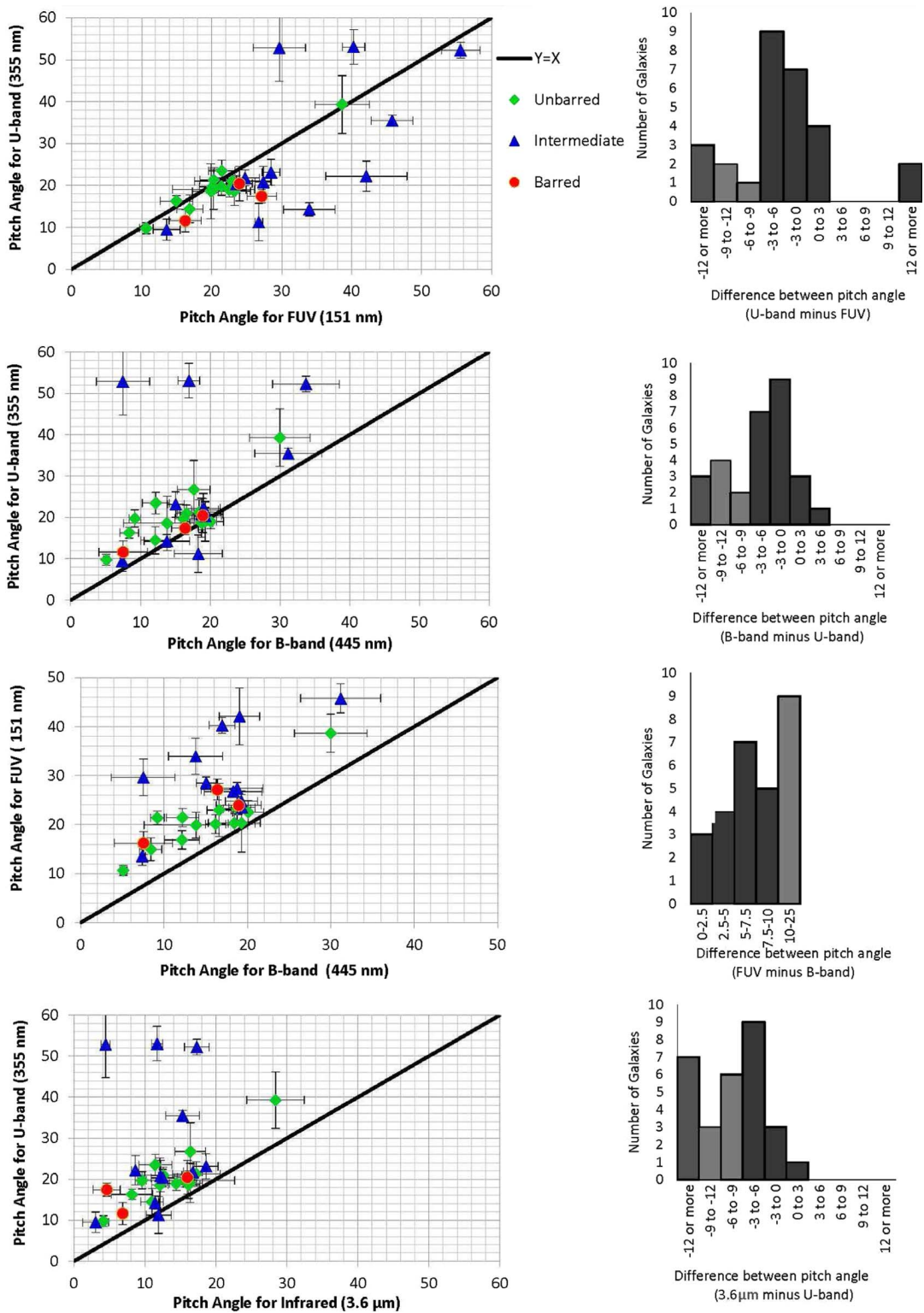

**Figure 4.** Comparisons between pitch angles measured at different wavelengths. Each point on the plots represents an individual galaxy positioned according to the measurement of its spiral-arm pitch angle at two different wavelengths. If pitch angle does not depend on wavelength, then galaxies should fall on the central diagonal line, indicating equality of pitch angle between the two wavelengths measured. The histograms accompanying each plot show the distribution of pitch angle differences (for those two wavelengths) in terms of the number of galaxies found in each bin. The histograms for the top plot shows that the *u* band and FUV wavelengths are fundamentally equal because the greatest numbers of galaxies have pitch angles at these wavelengths that agree to better than 3°. The same is true for the second plot, comparing *B*-band with *u*-band images. In contrast, we can see that FUV and *B*-band pitch angles (third plot) are different from each other, since in both cases the greatest numbers of galaxies have a pitch angle difference of more than 3° (see the relevant histogram), with very few found below 3° (this plot is reproduced from Pour-Imani et al. (2016). The same is true for *u*-band pitch angles and 3.6 μm pitch angles (bottom plot), which are clearly different from each other. Images of 29 galaxies were used at 355 nm, 445 nm, 3.6 μm, and 28 of these also had images at FUV (151 nm).

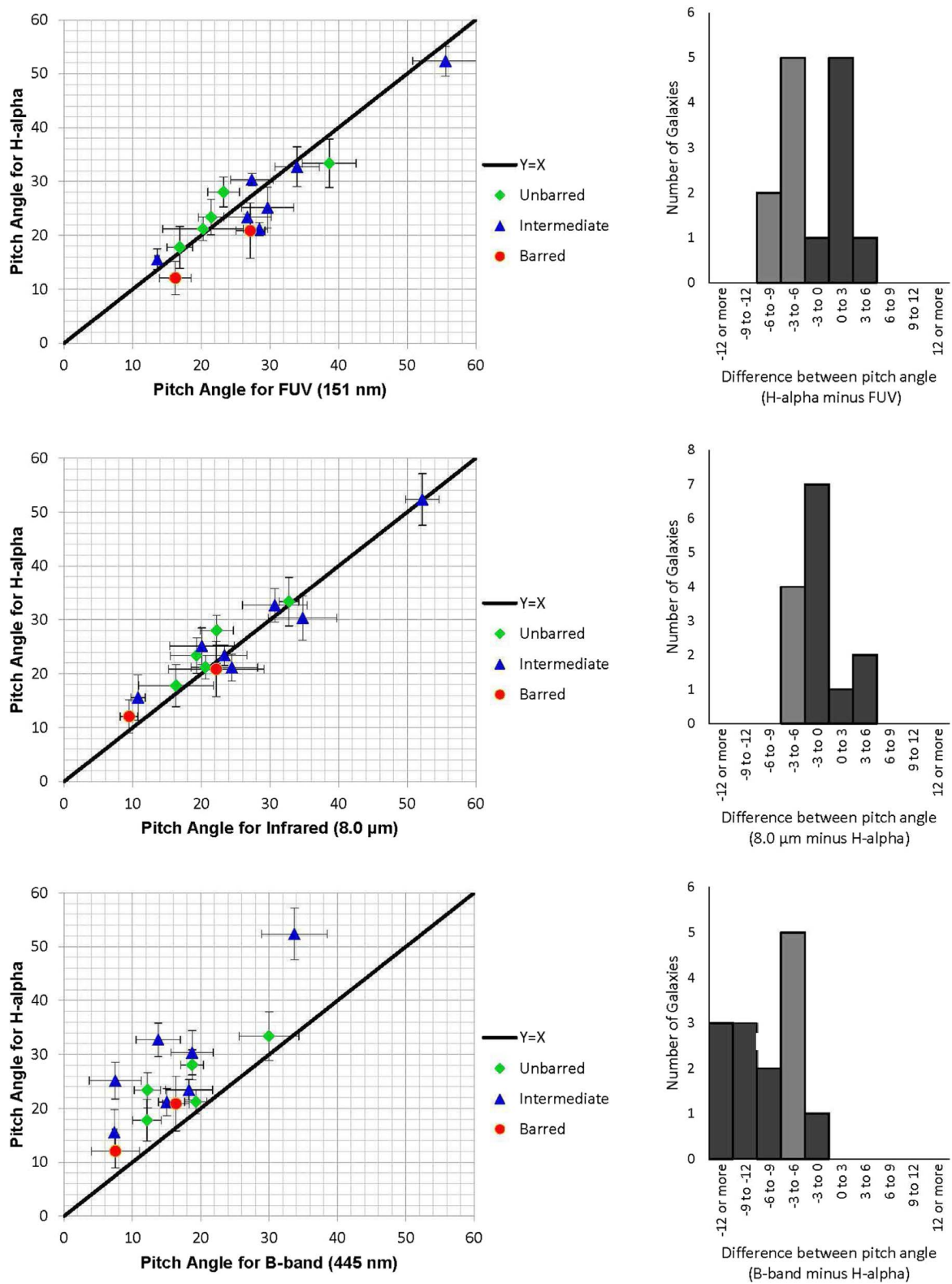

**Figure 5.** Comparisons between pitch angles measured at different wavelengths. Each point on the plots represents an individual galaxy positioned according to the measurement of its spiral-arm pitch angle at two different wavelengths. If pitch angle does not depend on wavelength then galaxies should fall on the central diagonal line indicating equality of pitch angle between the two wavelengths measured. The histograms accompanying each plot show the distribution of pitch angle differences (for those two wavelengths) in terms of the number of galaxies found in each bin. The histograms for the top plot shows that the H-$\alpha$ and FUV wavelengths are fundamentally equal because the greatest number of galaxies have pitch angles at these wavelengths that agree to better than 3°. The same is true for the second plot, comparing H-$\alpha$ with 8.0 $\mu$m images. In contrast, we can see that H-$\alpha$ pitch angles and *B*-band pitch angles (bottom plot) are different from each other, since in both cases the greatest number of galaxies have a pitch angle difference of more than 5° (see the relevant histogram), with very few found below 5°. Images of 14 galaxies were used at H-$\alpha$, 445 nm, 8.0 $\mu$m.

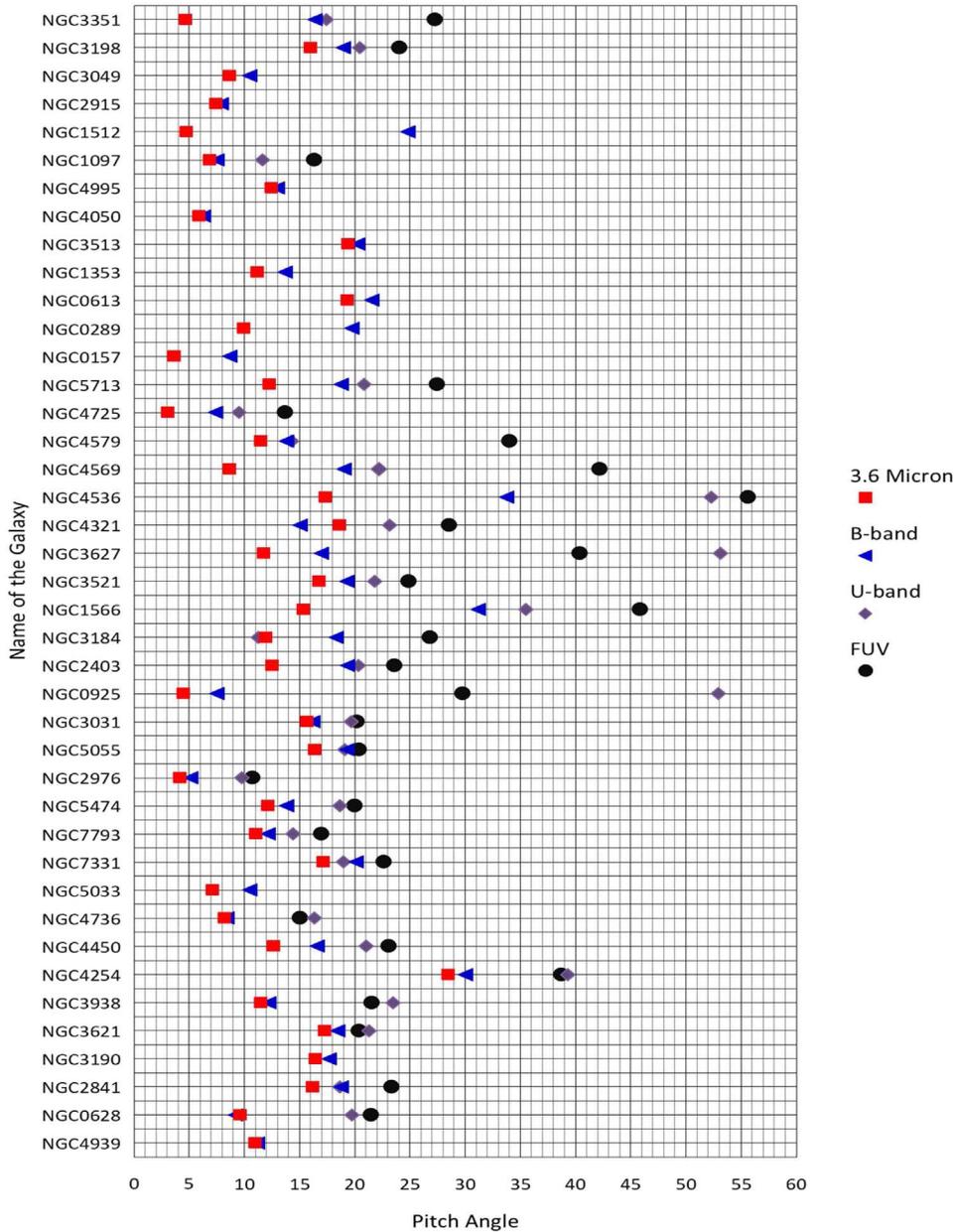

**Figure 6.** Evolution of measured pitch angle at different wavelengths from FUV (151 nm) to 3.6 μm for each individual galaxy.

### 4. The Location of the Density Wave

A careful analysis of our pitch angle measurements at different wavelengths suggests an overall picture very compatible with the density wave theory, but with one or two aspects that require closer examination.

The concise summary of our results is that examination of "stellar light" (optical and near-infrared wavelengths) produces pitch angles that are consistently tighter than those produced by wavelengths associated with the star-forming region. The wavelengths of light associated with the star-forming region include 8.0 μm, infrared light produced by warmed dust from clouds undergoing gravitational collapse, light emitted at the frequency of the H-α line produced by hot gas heated by protostars forming within the gas clouds, and FUV light at 151 nm emitted by very bright young stars. The rationale for this last assertion would be that the stars that emit most strongly in the FUV are sufficiently short-lived (O-type stars, for example) that they do not have time to move far from the star-forming region before they die. These stars simply form more quickly and live more briefly than other stars. Their great luminosity also increases the chance of their being seen while young, in spite of extinction. We see little in the way of measurable pitch angle differences within these two complexes. That is to say, the pitch angles measured in the FUV, H-α, and 8.0 μm wavebands all seem to be more or less the same (see Figures 4–6). Theoretically, there is an evolutionary relationship between these wavebands, with the order running from 8.0 μm initially, then H-α, then FUV (tracking initially warmed dust, followed by clouds heated from within by young stars and protostars, ending with very bright newly formed O-type stars). However, the truth is that this process takes place in a very short span of time, at most a few million years for the most luminous stars, and one should not expect to see a measurable difference in pitch angle between these different wavebands.

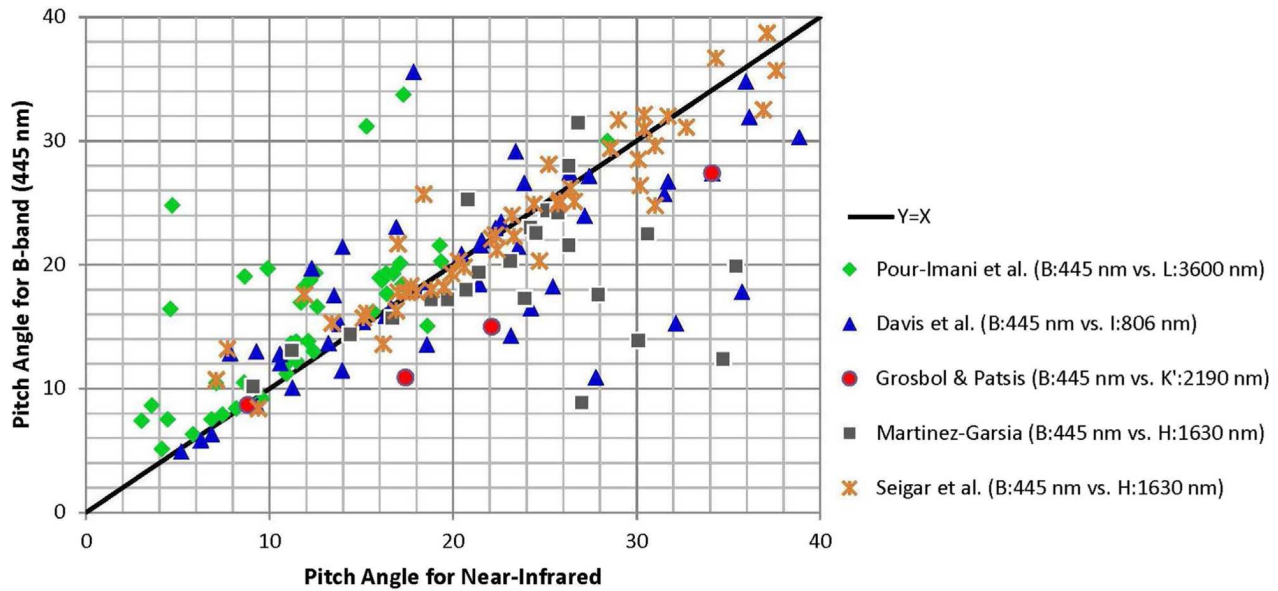

**Figure 7.** Comparison between pitch angles measured at the *B* band (445 nm) and the NIR. *B* band vs. L:3600 nm (Pour-Imani et al. 2016), I:806 nm (Davis et al. 2012), K′:2190 nm (Grosbol & Patsis 1998), H:1630 nm (Seigar et al. 2006; Martínez-García 2012).

Meanwhile, whatever differences exist between pitch angles measured at different optical and near-infrared wavelengths (that is to say, light produced by stars) they are small compared to the clear difference separating these optical pitch angles from the pitch angles measured in wavebands associated with the star-forming region. The fact that we see a looser pitch angle associated with the star-forming region and a tighter pitch angle associated with stars is entirely consistent with the stars being seen downstream from the star-forming region. The usual prediction of the density wave theory is that star formation occurs close to the position of the density wave (which compresses clouds of gas and sparks star formation) and that new stars, which may take some time to form and emerge from the clouds of gas and dust characteristic of the star-forming region, should be visible downstream from the density wave. Inside the corotation radius downstream means ahead of the density wave and outside the corotation radius it means behind (see Figure 1), so in pitch-angle terms, downstream means a lower pitch angle. At all optical and near-infrared wavelengths, we measure a tighter pitch angle than we do for star-forming wavelengths. Therefore, all our stellar light is "downstream" of the star-forming region.

There are, however, two regions where density-wave theory predicts an enhancement of stellar light. One is the current position of the density wave itself. At that location the density wave may compress not only clouds of gas, but also the distances between stars so that the starlight from that region is brighter than that from other parts of the disk. On the other hand, the density-wave theory has always claimed that our eye picks out the spiral pattern principally via the position of new stars. Star formation takes place near the location of the density wave and then, downstream of that position, we expect to see starlight enhanced by the turning on of young stars. This could be true not only in the *B* band but also into the NIR. Stellar evolution models certainly suggest that a great deal of red light is produced by young stars (see Figure 3). Therefore, the NIR light should also be enhanced downstream from the star-forming region.

To summarize, we argue that we are seeing a star-forming region close to the position of the density wave. Within the star-forming region, we see those stars that live less than 10 million years. They cannot move far enough to be seen downstream from where they were born. Downstream from the star-forming region we see light from recently born stars visible in the near-UV, optical, and NIR. Whether the density wave itself is also found close to this region of new stars will be the subject of further investigation. In a separate work (M. S. Abdeen et al. 2019, in preparation) we use rotation-curve data to show that the typical time elapsed in moving from, say, the FUV spiral arm to the *B*-band spiral arm, is on the order of 50 million years. This is certainly consistent with idea that we are seeing stars that were recently born, since many bright blue stars can be expected to live on the order of 100 million years.

## 5. Pitch Angle Differences between Optical and NIR Wavelengths

Looking at the optical and near-infrared wavebands, we do see some evidence of a pitch angle difference from the *B* band to 3.6 μm in the NIR. In our sample there is a tendency for them to be clustered to one side of the line of pitch angle equality, with the *B*-band pitch angles being slightly looser on average. However, the histogram in Figure 4 shows that the differences measured between the pitch angles at these two wavelengths are typically less than the average measurement error. There is a conflict here with some other experimental results. One theoretical scenario is that 3.6 μm would be sensitive to old red disk stars and thus should see a spiral arm at the current position of the density wave itself. In most scenarios, this would be the most upstream position and therefore the loosest pitch angle of all. Yet we find it to be perhaps the tightest of all. Some previous observers (Grosbol & Patsis 1998; Martínez-García 2012) have indeed found NIR waveband pitch angles to be looser than *B*-band pitch angles. We do not see this (see Figure 7). While Grosbol & Patsis (1998) reported a significant increase in pitch angle in the *K′* band over the *B* band, our results for the same galaxies show that the difference in *B*-band and *K′*-band pitch angle is different than the one quoted in Grosbol & Patsis (1998). Notably, this difference is much smaller than the difference reported by Grosbol & Patsis (1998; see Figure 8). In order to get the best possible sense of how things stand, let us carefully examine all of the available observational evidence. Two previous papers insisted that there is little or no variation in pitch angle across optical and NIR wavelengths (Seigar et al. 2006; Davis et al. 2012). We share the overall assessment of these papers in that we find that any difference between optical and NIR wavelengths is smaller than that of our typical measurement error.

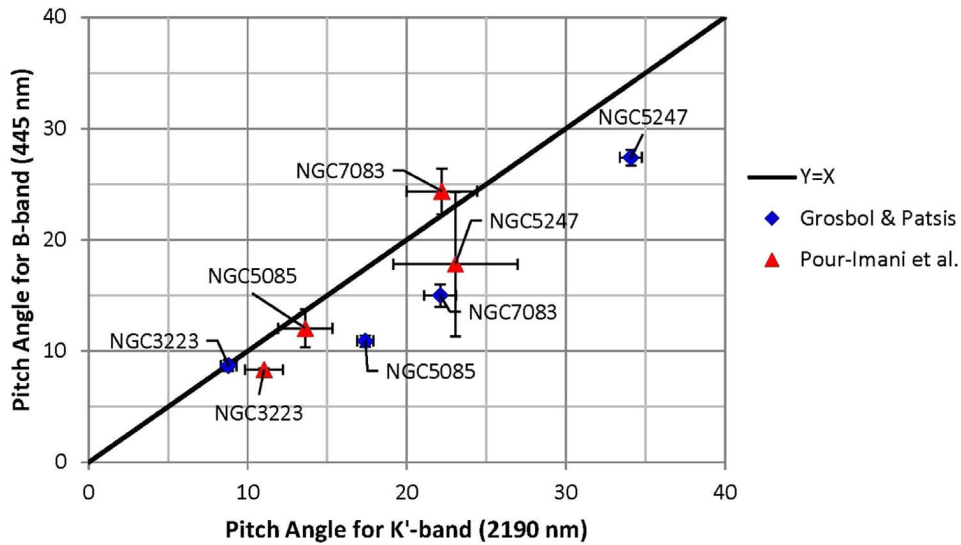

**Figure 8.** Comparison between our results and Grosbol & Patsis (1998) for the four galaxies (NGC 3223, 5085, 5247, 7083). The pitch angles are measured at the *B* band (445 nm) and the *K′* band (2190 nm).

We do see a small tendency for the 3.6 $\mu$m (redder) waveband to be tighter than the *B* band. This differs from the work of Grosbol & Patsis (1998), Martínez-García (2012), and Martínez-García et al. (2014). However, we first of all draw attention to the fact that in one of these papers (Martínez-García 2012) the result is arguably compatible with the results of Davis et al. (2012) and Seigar et al. (2006). Most of the points in Figure 11 of that paper are actually close to the line of equality. Only a minority are noticeably distant from it. The same is true of our data. So, looked at from a different point of view, most of the results to date are broadly compatible with the assertion that there is little variation of pitch angle across the optical or NIR wavebands. Martínez-García et al. (2014) and Grosbol & Patsis do not agree with this, but their samples are small (only five galaxies each) and smaller still when one considers that in both cases one or two of their galaxies do not follow the overall trend they report. So, it is important to state that there is no conclusive evidence of a decisive trend in pitch angle between the optical and NIR wavebands. More work is clearly needed before a definite conclusion can be drawn. In Figure 7, we have included on the same plot all the data from Grosbol & Patsis (1998), Seigar et al. (2006), Davis et al. (2012), Martínez-García (2012), and Pour-Imani et al. (2016). Each of these give pitch angle measurements in the *B* band and in a near-infrared band (which one varies with the study). It seems to us that this combined plot is consistent with the conclusion that there is no significant difference between the *B* band and near-infrared wavelength pitch angle measurements. How can this result be interpreted theoretically?

In addition to the two models already discussed (that the red light is due to young stars or to old red disk stars, or a mixture of both) is the possibility that a good deal of the NIR light could be coming from red supergiants. Obviously the fact that redder light is seen in the vicinity of the blue light raises the possibility that one is seeing the same young stars, some of which have reached the end of their lives. James & Seigar (1999) argued against this possibility but it could be a contributing factor to the red light being seen just slightly downstream from the blue light.

### 6. Evidence from the *u* band

One way of investigating further is to examine the pitch angles in the *u* band. This waveband falls between the *B* band (which is part of the stellar complex) and the FUV (which is part of the star formation complex). It is hard to imagine a scenario in which the FUV emission is not due to newly born massive stars. If the *B*-band light is from stars born in the same burst of star formation, seen a little further downstream, then we might expect the *u*-band pitch angle to fall between these two values.

We measured pitch angles for our sample in the *u* band and the results are given in Figure 4. We see that, predictably, the *u*-band pitch angle is not very different from either the FUV or the *B* band. But it is arguably true that it tends to be a little tighter than the FUV pitch angle and a little looser than the *B*-band pitch angle. Since there is a clear difference between the FUV and *B*-band pitch angles (see the histogram in the third panel of Figure 4), the fact that the *u* band manages to be close to both, even though they are not close to each other, further suggests that the *u* band falls somewhere between them. But if this suggests that the *B*-band light is indeed coming largely from newborn stars, then what should we believe about the 3.6 $\mu$m light? Elmegreen et al. (1989) argue that *B*-band and NIR spiral arms coincide and have a common origin, since their amplitudes are equal.

Admittedly, other observers have claimed that the amplitudes are not equal. Since we wish to point out the possibility that the light from new stars is responsible for the spiral arm in the NIR as well as the optical, let us discuss how this might be. Figure 3 (Figure 5 on page 35 of Eufrasio 2015) gives the spectrum of a cluster formed by a single burst of star formation at different stages of its evolution. The three relevant curves to note are the blue curve showing the spectrum of light from the cluster at age 10 Myr, the green curve for its spectrum at 100 Myr, and the yellow curve for one at a billion years. At any given position the luminosity at a given wavelength is essentially a sum of the new stars (10 Myr or less for the star formation pitch angle, and up to 100 Myr for the stellar pitch angle) plus the background of old disk stars (the 1 Gyr line or more).

At 151 nm (FUV) we see that there is a very high contribution from the new stars at 10m years and an essentially zero contribution from background stars. There is a sharp drop in the contribution from stars 100 Myr stars, so it is not surprising that at 151 nm we see the stars in the position of the star formation region. By the time 100 Myr has passed the luminosity has greatly decreased. Note that the speed at which stars move out of the star-forming region is slow because it is the relative speed of the star to the pattern speed that counts, which is typically some 20 or 30 pc/Myr. Looking at the $u$ band, which is at 365 nm, we see that the gap from 10 Myr to 100 Myr is a little less, but actually only a little.

The background contribution is more significant at the 1000 Myr line but it still significantly decreased in luminosity from the younger stars. Certainly, this suggests that the $u$-band spiral arm might not be too dissimilar from the FUV arm and this is what we see. Now, looking at the $B$ band (445 nm) we see that there is a noticeably smaller gap between the 10 Myr and 100 Myr curves, because of the bump in the 100 Myr curve that falls between the $u$ band and $B$ band.

The wavelength difference between the $u$ band and $B$ band is not great, but there is still a difference in the reduction from the 10 Myr curve to the 100 Myr curve. Of course, it is not a huge difference, which is not surprising given how close the two wavelengths are. Therefore, one might also expect that there would not be a big difference between the $u$ band and $B$ band pitch angles and this is again what we see. Since the $u$ band is close to both the FUV and $B$ band, but these two are distinguishable from each other, this suggests that, on average, $u$ band pitch angles are a little tighter than the FUV but a little looser than the $B$ band, as suggested by Figure 4. Since the $B$ band is part of the optical complex of pitch angles and FUV is part of the star-forming complex, it is in the $u$ band that we see a wavelength that falls between these two. One is tempted to see an evolution here (which would require further scrutiny and more data) from the FUV to the $u$ band to the $B$ band to the NIR, with each step being a small decrease in pitch angle. Each step in this sequence is too small for the difference in pitch angle to be greater than our average measurement error (between two and three degrees for pitch angles in this sample), but a double step typically shows a large enough difference to be greater than our measurement errors (as shown by the histograms in Figures 4 and 5). Thus, there is a decisive difference between the FUV and $B$ band and between the $u$ band and 3.6 $\mu$m. This tends to strengthen our belief that what we are seeing is a loose pitch angle in the star-forming spiral arm, which is most likely close to where the density wave lies, and then tighter and tighter pitch angles as one moves redward of the FUV.

Turning to the 3.6 $\mu$m images, the expectation has generally been that here one is sensitive primarily to the old red stellar population in the disk. These older disk stars have orbited the galaxy more than once and no longer display any enhanced density associated with clustering, because they have long since moved out of their original open clusters. However, when they pass through the density wave they presumably are compressed somewhat closer together by it. However, as mentioned already, this is not the only place where one expects to find a higher density of stars. Downstream from the star-forming region sparked by the density wave, the old disk population is augmented by a population of young stars. Looking at Figure 3, we see that this younger population produces plenty of red light. So, in fact, it is not out of the question that this blue-to-red gradient downstream from the star-forming region should be visible even in red or infrared light, through a combination of old red disk stars and younger stars.

In this context, it is worth noting that a few galaxies in Figure 4, all barred, have very large changes in pitch angle that are hard to reconcile with our scenario. These anomalies could be due to difficulties in measurement such as, close to the edge-on disk orientation, intrinsically flocculent spiral structure or the arc length of their spiral segments are short (pitch angles of these galaxies all have large error bars). Also, the pitch angles can be largely different when measured from different images with different qualities (Graham et al. 2019). Looking at Figure 4, we see that our outliers are not greater than the outliers from other studies. Increasing the sample size may help in identifying the reason for these odd results.

## 7. Conclusion

Our approach to testing density-wave theory is to look at the entire logarithmic spiral arm for differences in pitch angle. A different approach to investigating the same issue is to look for tracers of star formation and older stars in individual patches of the spiral arm. Recent studies of this type, such as Foyle et al. 2011 and Ferreras et al. (2012), report no consistent trend in positional offsets between these tracers. Thus, unlike us, they do not find evidence for this prediction of the modal density wave theory, and report their results as favoring the picture of density waves as transient structures that do not persist long enough to produce consistent offsets. Similar conclusions are drawn by those who use the radial Tremaine–Weinberg method to find results in support of the claim that the pattern speed of the spiral arm is actually radial-dependent (Merrifield et al. 2006; Meidt et al. 2009; Speights & Westpfahl 2011). It is true that comparing $B$-band with NIR images ,the pitch-angle measurements, including the work of several different groups as in Figure 7, show no overall pitch angle difference. This is certainly a point in favor of the transient picture. However, our results are not consistent with this picture regarding the consistent difference we see between pitch angles in the star-formation-tracing wavebands versus the stellar wavebands. Of course, we must recall that it has been proposed that different mechanisms for the formation of spiral arms operate in different galaxies. It is certainly possible that our sample favors grand-design spirals, with clearly logarithmic spirals and that as our sample is broadened, differences in behavior between individual galaxies will become more apparent. It may be true that some galaxies have transient spirals and others have more long-lived ones.

We have presented clear evidence in favor of a prediction of the density-wave theory that there is a tighter pitch angle in the $B$ band then is found for wavelengths associated with the star-forming region.

We have built upon the result in our earlier paper (Pour-Imani et al. 2016) by adding one more star-forming wavelength, the H-$\alpha$, which fits the pattern established by the 8.0 $\mu$m and FUV wavelengths from the earlier paper. We have also found that the $u$-band light seems to show pitch angles tending to fall between those of the star-forming region and the $B$ band, consistent with the standard interpretation of the density-wave theory. We regard our results as strongly in favor of the density-wave theory, though more work is certainly needed to reconcile our results with those reported from other methods. We do find evidence to support those earlier works which see only small differences in pitch angle between optical and NIR wavelengths. Broadly, we see two different systems, a star-forming region visible in the 8.0 $\mu$m, H-$\alpha$, and FUV and stellar light downstream visible in the $B$ band and the NIR. The $u$-band light, predictably, falls between these two systems. Thus we argue that we are seeing the star-forming region and then light from recently born stars moving downstream from that position. This is consistent with the long-lived density-wave theory of spiral structure.

The authors thank Bret Lehmer and Jerry Sellwood for valuable suggestions that guided our research. This research has made use of the NASA/IPAC Extragalactic Database, which is operated by the Jet Propulsion Laboratory, California Institute of Technology, under contract with the National Aeronautics and Space Administration.